\documentclass[a4paper,11pt]{article}
\pdfoutput=1
\usepackage{jheppub}
\usepackage[T1]{fontenc}
\def\ee{\end{equation}}
\def\bea{\begin{eqnarray}}
\def\eea{\end{eqnarray}}

\usepackage{xcolor}
\def\lsim{\raise0.3ex\hbox{$\;<$\kern-0.75em\raise-1.1ex\hbox{$\sim\;$}}}
\def\gsim{\raise0.3ex\hbox{$\;>$\kern-0.75em\raise-1.1ex\hbox{$\sim\;$}}}

\usepackage{epsfig}
\usepackage{amsmath}
\usepackage{amsfonts}
\usepackage{float}
\usepackage{amssymb}
\usepackage{xcolor}
\usepackage{amssymb}
\def\be{\begin{equation}}
\def\ee{\end{equation}}
\def\bea{\begin{eqnarray}}
\def\eea{\end{eqnarray}}

\DeclareMathOperator{\diag}{diag}

\title{\boldmath Constraints on the $X17$ boson from IceCube searches for non-standard  interactions of neutrinos}

\author[a]{Rikard Enberg,}
\author[b,c]{ Ya\c{s}ar Hi\c{c}y\i lmaz,}
\author[a,c]{ Stefano Moretti,}
\author[a]{ Carlos P\'erez de los Heros}
\author[a]{ and  Harri Waltari}

\affiliation[a]{Department of Physics \& Astronomy, Uppsala University, Box 516, SE-751 20 Uppsala, Sweden}
\affiliation[b]{Department of Physics, Bal\i kesir University, TR10145, Bal\i kesir, Turkey}
\affiliation[c]{School of Physics and Astronomy, University of Southampton, Highfield, Southampton SO17 1BJ, United Kingdom.}

\abstract{
We explain the ATOMKI anomaly with a very light $Z'$ state that features non-anomalous and non-flavour-universal vector and axial-vector couplings to all leptons. This $Z'$ comes from a theoretical framework with a spontaneously broken $U(1)'$ 
symmetry in addition to the Standard Model gauge group and is compliant with current measurements of the
anomalous magnetic moments of the electron and the muon as well as beam dump experiments. The lepton flavour structure of this model 
allows for $Z'$ couplings to all light neutrinos, suggesting the possibility of $Z'$-mediated Non-Standard Interactions (NSIs) of neutrinos in matter, so that measurements of the strength parameters of the NSIs can constrain the value of the couplings. We use experimental constraints on NSIs of neutrinos using older TEXONO data and newer IceCube data. The IceCube data, in particular, strongly constrain the flavour universality of the leptonic vector current.
The constraints enable us to define the region of parameter space of this theoretical scenario that can be pursued in further phenomenological analyses.}

\begin{document}

\maketitle \flushbottom

\section{Introduction}
\label{sec:intro}

The discovery of the Higgs boson~\cite{CMS:2012qbp,ATLAS:2012yve} provided strong evidence that the Standard Model (SM) of particle physics is indeed a consistent, and successful, description of elementary particles and their interactions, at least at the energies probed so far in accelerators. There are, however, several experimental ``anomalies'' that could point to new physics Beyond the SM (BSM). The majority of the experimental results that cannot be explained within the SM have been uncovered in non-LHC experiments, such as $(g-2)_\mu$, the measured value of the magnetic moment of muon in the Muon $g-2$ experiment at Brookhaven National Laboratory (BNL) \cite{Muong-2:2006rrc}. Another anomaly is the significant enhancement of more than 5$\sigma$ in the invariant mass and angular distributions of electron-positron final states of decays of excited $^{8}$Be measured by the ATOMKI collaboration in 2015~\cite{Krasznahorkay:2015iga}. Most studies trying to understand this result have demonstrated that standard nuclear physics or QCD cannot lead to a satisfactory explanation~\cite{Alves:2023ree,Zhang:2017zap,Koch:2020ouk,Hayes:2021hin,Viviani:2021stx}. The ATOMKI results can be accounted for by the existence of a new vector or axial-vector mediator with a mass of around 17 MeV, which has been called the $X17$ boson~\cite{Feng:2016jff,Feng:2016ysn,Feng:2020mbt,Nomura:2020kcw,Seto:2020jal,Kozaczuk:2016nma,DelleRose:2018pgm,Nam:2019osu,Zhang:2020ukq}. Further studies~\cite{Barducci:2022lqd,Denton:2023gat} demonstrated that, due to conflicts with the non-observation of deviations from the SM in neutrino scattering experiments,  the scenarios including a pure vector  mediator  are less favourable,  while an axial-vector state appears as the most promising candidate to simultaneously explain all the anomalous nuclear decays reported by the ATOMKI collaboration~\cite{Alves:2023ree}. In the case of a pure vector boson, the couplings of new boson to nucleons are strongly constrained by the NA48/2 experiment~\cite{NA482:2015wmo}. Moreover, the pure vector boson explanations could be ruled out under the strong experimental limits on pion decays~\cite{Hostert:2023tkg}, such as SINDRUM~\cite{SINDRUM:1989qan}. 

As a minimal approach, a family-dependent $U(1)'$ extension of the SM would be an ideal way to allow axial-vector couplings that could explain the ATOMKI anomaly while the vector couplings are also present. In this framework, the Yukawa interactions are modified by higher-dimensional operators~\cite{Pulice:2019xel,DelleRose:2018eic}. This scenario, which introduces a new light vector boson, $Z'$, also leads to  Non-Standard Interactions (NSIs) of neutrinos  that affect neutrino flavour ratios  in matter \cite{Proceedings:2019qno}. Currently, limits from the TEXONO experiment have been derived on the combination $\sqrt{\epsilon_{e}\epsilon_{\nu}}$, where $\epsilon_{e}$ and $\epsilon_{\nu}$ are the couplings of the $Z^{\prime}$ to  electrons and neutrinos, respectively. The limits imply that $\sqrt{\epsilon_{e}\epsilon_{\nu}} < 7\times 10^{-5}$ for constructive interference and $\sqrt{\epsilon_{e}\epsilon_{\nu}} < 3\times 10^{-4}$ for destructive interference \cite{TEXONO:2009knm,Bilmis:2015lja,Feng:2016ysn}.
Moreover, the experimental constraints on NSIs from neutrino oscillations can be applied to restrict the 
couplings of the new boson with SM fermions.  For instance, in Ref.\ \cite{Cabarcas:2023dkj}, the authors presented the experimental NSI constraints on lepton flavour violating couplings for a $Z'$ boson with a mass heavier than the $ \tau $ lepton. In this paper, we will confront the ATOMKI anomaly and the anomalous magnetic moments of leptons with the NSI constraints reported by the IceCube collaboration~\cite{IceCubeCollaboration:2021euf}. We will show the allowed regions of couplings and the amount of  non-universality in the minimal $U(1)'$ extension of the SM which satisfy IceCube constraints.

The rest of the paper is organised as follows. In Sec.~\ref{sec:model} we provide a brief discussion of the main components of the model. We discuss the general formalism of NSI dynamics in Sec.~\ref{sec:nsi} and  the new constraints from IceCube that we use in our analysis in Sec.~\ref{sec:external}. After summarising our computational procedure and enforcing experimental constraints in Sec. \ref{sec:numeric}, we present our results over the surviving parameter space of couplings and NSI parameters in Sec.~\ref{sec:results}. Finally, we summarise and conclude in Sec.~\ref{sec:conc}.

\section{The $U(1)'$ model}
\label{sec:model}
We focus on an extension of the SM with a generic $U(1)'$  symmetry which mixes with the SM $U(1)_Y$. The kinetic term of the Lagrangian is given by
\begin{equation}
\label{eq:KineticL}
\mathcal{L}_\mathrm{Kin}^{U(1)'} = - \frac{1}{4} \hat F_{\mu\nu} \hat F^{\mu\nu}  - \frac{1}{4} \hat F'_{\mu\nu} \hat F^{'\mu\nu} - \frac{\eta}{2} \hat F'_{\mu\nu} \hat F^{\mu\nu},
\end{equation}
where $ F_{\mu\nu} $  and $ F'_{\mu\nu} $  are the field strengths of the gauge fields $ B_\mu $ and $ B'_\mu$ that correspond to $U(1)_Y$ and $U(1)'$, respectively, and $\eta$ quantifies the kinetic mixing between these abelian symmetries. The  Lagrangian for the gauge sector can be diagonalised via a GL$(2,\mathbb{R})$ transformation which rescales the gauge fields as 
\begin{eqnarray}
\left(
\begin{array}{c}
\hat{B}_{\mu} \\	\hat{B}'_{\mu}
\end{array}
\right) 
&=&	
\left(
\begin{array}{cc}
1	&	-\frac{\eta}{\sqrt{1-\eta^2}}	\\
0	&	\frac{1}{\sqrt{1-\eta^2}}
\end{array}
\right)
\left(
\begin{array}{c}
B_{\mu} \\	B'_{\mu}
\end{array}
\right)	\, ,
\label{eq:kinZZprotation}
\end{eqnarray}
where $ \hat{B}_{\mu} $ and $ \hat{B}'_{\mu} $ are original $ U(1)_{Y} $ and $ U(1)' $ gauge fields with off-diagonal kinetic terms while $ B_{\mu} $ and $ B'_{\mu} $ do not possess such terms. After the transformation in Eq. (\ref{eq:kinZZprotation}), the gauge covariant derivative can be written as
\begin{equation}
\label{CovDer}
{\cal D}_\mu = \partial_\mu + \dots + i g_1 Y B_\mu + i (\tilde{g} Y + g' Q') B'_\mu, 
\end{equation}
where $Y$ and $g_1$ are the hypercharge and its gauge coupling while $Q'$ and $g'$ in the additional term are the $U(1)'$ charge and its gauge coupling, respectively. In addition, $\tilde{g}$ is the mixed gauge coupling between the two gauge groups. In Eq.~(\ref{CovDer}), we have introduced  the $g'$ and $\tilde{g}$  in terms of original diagonal gauge couplings and the kinetic mixing parameter $ \eta $ as

\begin{align}
	\label{tan-chi}
	{g'} \equiv \frac{{g}_{U(1)'}}{\sqrt{1-\eta^2}},~~ \: \: \tilde{g} \equiv -\frac{\eta g_1}{\sqrt{1-\eta^2}}.
\end{align}

The scalar potential of the model is 
\begin{eqnarray}
\label{eq:HM}
V(H,\chi) &=&   -\mu^2|H|^2 +  \lambda |H|^4 -\mu_\chi^2 |\chi|^2 + \lambda_\chi |\chi|^4 \\
&+&\kappa  |\chi|^2|H|^2\, , \nonumber
\end{eqnarray}
where $H$ is the SM Higgs doublet and $\kappa$ is the mixing parameter that connects the SM and $ \chi $ Higgs fields. 
Unlike the SM Higgs sector, there are two physical Higgs states with their  acquired Vacuum Expectation Values (VEVs) 
\bea
\label{eq:vev}
\langle H \rangle = \frac{1}{\sqrt{2}} \left( \begin{tabular}{c} 0 \\ $v$ \end{tabular} \right) \,, \qquad \langle\chi\rangle = \frac{v'}{\sqrt{2}}\,.
\eea

After Spontaneous Symmetry Breaking (SSB),  the relation of  the two CP-even mass eigenstates $h_1$ and $h_2$ with these physical  states via an orthogonal transformation through a  mixing matrix can be written as

{\bea
\left( \begin{array}{c} h_1 \\ h_2 \end{array} \right) = \left( \begin{array}{cc} \cos \theta & - \sin \theta \\  \sin \theta & \cos \theta \end{array} \right)  \left( \begin{array}{c} H  \\ \chi \end{array} \right),
\eea 
where the mixing angle $\theta$ varies over the interval  $- \pi/2 < \theta < \pi/2$. The mass eigenstates are
\bea
m_{h_{1,2}}^2 = \lambda v^2 + \lambda\chi  v'^2 \mp \sqrt{\left( \lambda v^2 - \lambda\chi  v'^2\right)^2 + \left( \kappa v v' \right)^2} \,,
\eea
from where, using the the physical masses $m_{h_{1,2}}$, the two VEVs $v, v'$ and with the initial conditions of  scalar couplings,  one can get the following relation for the mixing angle  \cite{Coriano:2015sea}
\bea
\tan 2 \theta = \frac{\kappa v v'}{\lambda v^2 - \lambda\chi  v'^2}.
\label{theta}
\eea}
{In this work, we assume that $ h_2 $ is dominantly the SM-like Higgs boson while the exotic boson $ h_1 $ is dominantly a singlet-like  Higgs state. In Ref.\ \cite{Hicyilmaz:2022owb}, the possible $ Z' $ signatures mediated by such Higgs bosons were worked out.}

The $U(1)'$ symmetry is broken by a new scalar $ \chi $, which is a singlet under the SM gauge group and has $U(1)'$ charge $Q'_\chi$ and its VEV   shown in Eq. ~(\ref{eq:vev}). The spontaneous breaking of the $U(1)'$ symmetry results in the mass term of a new vector boson. The mass eigenstates of the neutral gauge bosons are obtained by the transformation as
\bea
\left( \begin{array}{c} B^\mu \\ W_3^\mu \\ B'^\mu \end{array} \right) = \left( \begin{array}{ccc} 
	\cos \theta_W & - \sin \theta_W \cos \theta' & \sin \theta_W \sin \theta' \\
	\sin \theta_W & \cos \theta_W \cos \theta' & - \cos \theta_W \sin \theta' \\
	0 & \sin \theta'  & \cos \theta'  
\end{array} \right)
\left( \begin{array}{c} A^\mu \\ Z^\mu \\ Z'^\mu \end{array} \right),
\eea
where $\theta_W$ is the usual Weinberg angle  and $\theta'$ is $Z-Z'$ mixing angle defined as \cite{Coriano:2015sea}
\bea \label{ThetaPrime}
\tan 2 \theta' = \frac{2 g_H \sqrt{g_2^2 + g_1^2}}{ g_H^2 + ( 2 Q'_\chi g' \, v'/v )^2 - g_2^2 - g_1^2} \,,
\eea
where  $ g_H = \tilde{g} + 2 g' Q'_H $. The mixing angle $\theta'$ is strictly constrained by the bounds from the precision tests in the LEP experiment  as $|\theta'| \lesssim 10^{-3}$ \cite{DELPHI:1994ufk,Erler:2009jh}. In the case of small $\theta'$, the gauge boson masses can be read in terms of $m_{Z,Z'}$ as
\bea \label{ZZpMassesSmallTh}
m_Z \simeq \frac{v}{2} \sqrt{g_2^2 + g^2}\,, \qquad m_{Z'} \simeq \frac{v}{2} \sqrt{g_H^2 + (2 Q'_\chi g' \, v'/v)^2} \,,
\eea
while $\theta'$ is rewritten as
\bea \label{ThetaPrimeMasses}
\theta' \simeq g_H \frac{ m_{Z}^2 }{m_{Z'}^2-m_{Z}^2} \,.
\eea
  In the case of $g' \sim {\cal O}(10^{-4} - 10^{-5})$, $M_{Z'}$ would be light, with a mass  of ${\cal O}(10)$ MeV, which is the desired mass region for a  potential solution of the ATOMKI anomaly if  $ v' $  is of order
$ {\cal O}(100 -1000) $ GeV.

The Lagrangian that describes the interactions of the extra gauge boson $  Z' $ with SM fermions is 
\begin{eqnarray}
\label{eq:NeuCurLag}
\mathcal{L}^\mathrm{Z'} &=& \bar q \gamma^\mu \left( C^{qq'}_{L} P_L + C^{qq'}_{R} P_R \right) q' Z'_\mu + \bar \nu_l \gamma^\mu \left( C^{ll'}_{L} P_L \right) \nu_{l'} Z'_\mu \nonumber \\ 
&+&\bar l \gamma^\mu \left( C^{ll'}_{L} P_L + C^{ll'}_{R} P_R \right) l' Z'_\mu,
\end{eqnarray}
where $ q^{(\prime)}$,  $l^{(\prime)}$  and $  \nu_{l^{(\prime)} }$ refer to up-type/down-type quarks, charged leptons and their neutrinos while $ C^{XX}_{L} $ and $ C^{XX}_{R} $ are Left ($L$) and Right ($R$) handed couplings and $P_L$ and $P_R$ the corresponding projection operators $\frac{1\mp\gamma^5}{2}$, respectively. In our model, there are no flavour-violating (non-diagonal) coupling terms for the quark and lepton sector while the flavour-conserving (diagonal) $ f=f' $ coupling terms are written as 
\begin{eqnarray}
	C^{ff}_{L} \!&\!=\!&\!  - g_Z \sin \theta' \left( T^3_f - \sin^2 \theta _W Q_f \right) \!+\! ( \tilde g Y_{f, L} \!+\! g' Q'_{f, L})  \cos \theta'\!, \label{CffL}  \\
	C^{ff}_{R} &=&  g_Z \sin^2 (\theta _W) \sin(\theta') Q_f + ( \tilde g Y_{f, R} + g' Q'_{f, R}) \, \cos \theta ', \label{CffR}
	\end{eqnarray}
where $g_Z = \sqrt{g_1^2 + g_2^2}$ is the Electro-Weak (EW) coupling.  Here,  $T_f ^3$ and $Q_f$  denote the weak isospin and electric charge of the fermion $f$, respectively. Finally, $Y_{f,L/R}$ and $Q'_{f,L/R}$ indicate the hypercharge and $U(1)'$ charges of the $L/R$-handed fermion.

Considering the interactions in Eq.~(\ref{eq:NeuCurLag}), the $ Z' $ boson undoubtedly contributes to the Anomalous Magnetic Moments (AMMs) of  the charged leptons $a_f$, for $ f= e, \mu, \tau$. Such contributions could be determined by their vector and axial couplings  and the mass of the $ Z' $ boson $ m_{Z'}  $ as  \cite{Leveille:1977rc}
\begin{eqnarray}
	\Delta a_\textit{f} &=& \frac{m_\textit{f}^2}{4\pi^2 m_{Z'}^2} \Big( C_{\textit{f}, V}^2 \int_0^1 \frac{x^2 (1-x)}{1 - x + x^2 m_\alpha^2/m_{Z'}^2} dx \nonumber\\
	&-& C_{\textit{f}, A}^2 \int_0^1 \frac{x (1-x) (4 - x) + 2 x^3 m_\textit{f}^2/ m_{Z'}^2}{1 - x + x^2 m_\textit{f}^2/m_{Z'}^2} dx \Big),
	\label{eq:gm2sfull}
\end{eqnarray}
where the vector and axial couplings defined by Eqs. (\ref{CffL}) and (\ref{CffR}) as

\bea \label{VAcouplings}
C_{f, V} = \frac{C^{ff}_{R} + C^{ff}_{L}}{2} \,,  C_{f, A} = \frac{C^{ff}_{R} - C^{ff}_{L}}{2} .
\eea

 In the limit of small gauge coupling and mixing, $ g^\prime, \tilde{g} \ll 1$, we can redefine the vector and axial couplings as \cite{DelleRose:2018pgm}
 
 \begin{align}
&C_{f, V} \approxeq \tilde{g} \cos^2 (\theta _W) Q_f  + g^\prime [Q'_H (T^3_f  - 2 \sin^2 (\theta _W) Q_f ) + (Q'_{f, R}+Q'_{f, L})/2]\, , \nonumber \\
&C_{f, A} \approxeq g^\prime [-Q'_H T^3_f + (Q'_{f, R}-Q'_{f, L})/2]\, .
 \end{align}
  For the limits $m_f \ll m_{Z'}$ and $m_f \gg m_{Z'}$, Eq.~(\ref{eq:gm2sfull}) reduces to \cite{Bodas:2021fsy}
\begin{equation}
	\Delta a_f \simeq \begin{cases}
		m_f^2 \left( C_{f, V}^2 - 5 C_{f, A}^2 \right)/(12\pi^2 m_{Z'}^2)~, & m_f \ll m_{Z'} \, , \\
		(m_{Z'}^2 C_{f, V}^2 - 2m_f^2 C_{f, A}^2)/(8\pi^2 m_{Z'}^2) ~, & m_f \gg m_{Z'} \, .
	\end{cases}
	\label{gm2s}
\end{equation}

For $M_{Z'} \simeq 17$ MeV, one then finds 
\bea \label{gm2s2}
\Delta a_{e} &=& 7.6 \times 10^{-6} C_{e,V}^2  -3.8 \times 10^{-5} C_{e,A}^2 ,\\
\Delta a_{\mu} &=&0.009 C_{\mu,V}^2 -C_{\mu,A}^2 .
\eea 

The contribution from $Z'$ to $B_{s}\rightarrow \mu^{+}\mu^{-}$ can be parameterised as \cite{Bobeth:2013uxa}
\begin{equation}
BR(B_{s}\rightarrow \mu^{+}\mu^{-})=\frac{|V_{tb}V_{ts}^{*}|^{2}G_{F}^{2}}{16\pi^{3}}m_{B_{s}}m_{\mu}^{2}\alpha^{2}\tau_{B_{s}}f_{B_{s}}^{2}\sqrt{1-\frac{4m_{\mu}^{2}}{m_{B_{s}^{2}}}}(C_{10\mu}-C_{10^{\prime}\mu})^{2}.
\end{equation}
For detailed information about Z' contribution to other rare B decays, please see in Ref. \cite{Buras:2012dp}.

Furthermore, one of the new physics effect of the $Z'$ boson is manifest  in neutrino scattering experiments in the presence of  non-zero neutrino couplings in Eq.~(\ref{eq:NeuCurLag}). If we consider the TEXONO experiment, which probes  neutrino-electron scatterings as mentioned in Sec. \ref{sec:intro}, the correction to the SM cross-section of anti-neutrino scattering for $ m_{Z'} \gsim 10 $ MeV  in terms of the Z'-electron couplings is \cite{Bodas:2021fsy,Lindner:2018kjo}
	\begin{align}
\frac{\sigma(\overline{\nu_e}  e^- \to \overline{\nu_e}  e^-)}{\sigma(\overline{\nu_e} e^- \to \overline{\nu_e}  e^-)^\text{\rm SM}} \simeq  1 & + \left(2.07 C_{e,V} + 1.39 C_{e,A} \right) 10^{11} C^{\nu_e \nu_e}_L \left(\frac{\text{MeV}}{m_{Z'}} \right)^2 \notag \\
& + \left( 1.37 C_{e,V}^2 + 2.62 C_{e,V} C_{e,A} + 1.64 C_{e,A}^2 \right) \left( 10^{11} C^{\nu_e \nu_e}_L \right)^2 \left( \frac{\text{MeV}}{m_{Z'}} \right)^4 ~.
\label{texono}
\end{align} 

It is easy to note that AMM and neutrino scattering  experiments put additional bounds on $Z'$-fermion couplings as well as neutrino oscillation experiments such as IceCube via NSI effective couplings.

Let us now consider the Yukawa sector of our model. The Yukawa Lagrangian of the SM is 
\be 
- \mathcal{L}_{\rm Yuk.}^{\rm SM} = Y_u \bar{Q} \tilde{H} u_R + Y_d \bar{Q} H d_R + Y_e \bar{L} H e_R.
\label{Yukawa}
\ee
In order to satisfy  gauge invariance, the charges of the fields under the $U(1)'$ group must obey the condition of
\begin{equation}
-Q'_Q -Q'_H +Q'_u = -Q'_Q + Q'_H +Q'_d = -Q'_L + Q'_H + Q'_e = 0.
\label{eq:gauge_invariance}
\end{equation}
Considering Eqs. (\ref{CffL}) and (\ref{CffR}) with these relations, one can find no axial-vector couplings to the $Z'$ for quarks and leptons. In this work, our theoretical model relies on flavour-dependent charges of the $Z'$. Having such non-universal $U(1)'$ charges allows axial-vector couplings of the $Z'$ with nucleons, which are crucial to overcome the strict experimental bounds in the pure vector coupling case \cite{Feng:2016jff,Feng:2016ysn,NA482:2015wmo}. To achieve this, a new mechanism is identified that generates masses and couplings of the first two fermion generations at higher orders as the SM-like Yukawa interactions are available only for the third generation. In this case, the Yukawa interaction terms for first two fermion families  are modified as
\bea
- \mathcal{L}_{\rm Yuk.} &=& \Gamma^{u} \dfrac{\chi^{n_{ij}}}{M^{n_{ij}}} \overline{Q}_{L,i}\tilde{H}u_{R,j} 
+ \Gamma^{d} \dfrac{\chi^{l_{ij}}}{M^{l_{ij}}} \overline{Q}_{L,i} H d_{R,j} \nonumber \\
&+&\Gamma^{e} \dfrac{\chi^{m_{ij}}}{M^{m_{ij}}} \overline{L}_{i} H e_{R,j}+ h.c.,
\eea
where $M$ is the non-renormalisable scale whose dimension is determined by the $U(1)'$ charges of the involved fields \cite{DelleRose:2018eic}. Here, we are able to obtain axial-vector couplings for the first two generations of quarks to successfully reproduce the ATOMKI anomaly,
\begin{align}
&-Q'_{Q_{1,2}} -Q'_H +Q'_{u_{1,2}} \neq 0 \nonumber \\
&-Q'_{Q_{1,2}} + Q'_H +Q'_{d_{1,2}} \neq 0.
\end{align}

The charges must also satisfy the anomaly cancellation conditions for the fermionic content of the SM and the additional $R$-handed neutrinos:
\begin{align}
\label{eq:anomaly}
& \sum_{i=1}^{3} (2 Q'_{Q_i} - Q'_{u_i} - Q'_{d_i}) = 0 \,,  \\
&  \sum_{i=1}^{3} \, ( 3 Q'_{Q_i} +  Q'_{L_i})  = 0 \,,  \\
& \sum_{i=1}^{3} \left( \frac{Q'_{Q_i}}{6} - \frac{4}{3} Q'_{u_i} - \frac{Q'_{d_i}}{3}  +   \frac{Q'_{{L_i}}}{2} - Q'_{e_i}\right) = 0 \,,  \\
& \sum_{i=1}^{3} \left( Q_{Q_i}^{\prime 2} - 2 Q_{u_i}^{\prime 2}  + Q_{d_i}^{\prime 2}       - Q_{{L_i}}^{\prime 2}  + Q_{e_i}^{\prime 2}  \right) = 0 \,,  \\
& \sum_{i=1}^{3} \left( 6 Q_{Q_i}^{\prime 3}  - 3 Q_{u_i}^{\prime 3}  - 3 Q_{d_i}^{\prime 3}  +  2 Q_{{L_i}}^{\prime 3}  - Q_{e_i}^{\prime 3} \right)  + \sum_{i=1}^{3} Q_{\nu _i}^{\prime 3}    = 0 \,,  \\
& \sum_{i=1}^{3} \left( 6 Q'_{Q_i} - 3 Q'_{u_i} - 3 Q'_{d_i}  + 2 Q'_{{L_i}} - Q'_{e_i} \right) + \sum_{i=1}^{3} Q'_{\nu _i}    = 0 .
\label{eq:anomaly2}
\end{align}
In addition to these conditions, we also impose that the first two generations of quarks are flavour-universal under  $U(1)'$ in order to alleviate experimental bounds on flavour violating interactions of quarks, i.e., $Q'_{Q_1} = Q'_{Q_2}  \,,  
 Q'_{d_1} = Q'_{d_2} \,,   
 Q'_{u_1} = Q'_{u_2}$.
Conversely, for the purpose of this study, the $ U(1)' $ charges of the  lepton sector were left as fully non-universal. We can then express the
fifteen charges in terms of the four charges $ Q'_{L_1},  Q'_{Q_1}, Q'_{Q_3}, Q'_{\nu_1}$ that are free model parameters,
by using the conditions  
\begin{align}
\label{eq:charges}
& Q'_{L_1}+Q'_{L_2} + 6 Q'_{Q_1} = 0  \,,  \nonumber \\
& Q'_{L_3} + 3 Q'_{Q_3} = 0  \,,  \nonumber \\
& Q'_{d_1}+ 2 Q'_{Q_1} = 0  \,,  \nonumber \\
& Q'_{d_3}+ 2 Q'_{Q_1} = 0  \,,  \nonumber \\
& Q'_{u_1} - 4 Q'_{Q_1} = 0  \,,  \nonumber \\
& Q'_{u_3}- 2 Q'_{Q_1}- 2 Q'_{Q_3} = 0  \,,\nonumber   \\
& Q'_{e_1} + Q'_{L_1} + 9 Q'_{Q_1}= 0 \,,  \nonumber \\
& Q'_{e_2} - Q'_{L_1} + 3 Q'_{Q_1} = 0 \,,  \nonumber \\
& Q'_{e_3} + 4 Q'_{Q_3} + 2 Q'_{Q_1} = 0  \,, \nonumber  \\
& Q'_{\nu_1}+Q'_{\nu_2} = 0  \,,  \nonumber \\
& Q'_{\nu_3} - 2 Q'_{Q_3} + 2 Q'_{Q_1}= 0  \,,  \nonumber \\
& Q'_{H}- 2 Q'_{Q_1} - Q'_{Q_3} = 0  \,.
\end{align}
The relations in Eq.~(\ref{eq:charges}) are derived from Eqs.~(\ref{eq:anomaly}--\ref{eq:anomaly2}) and from the Yukawa conditions for only the third fermion family in Eq.~(\ref{eq:gauge_invariance}), and allow us to find numerical solutions satisfying the anomaly cancellation equations exactly. We scan numerically over integer solutions for the free charges, for absolute values from 1 to $ Q_\text{max}=200 $. Larger $ Q_\text{max} $ is possible, but since the number of solutions grows rapidly with $  Q_\text{max} $, this can lead to numerical problems. We then normalise all integer solutions by dividing by 100, thus obtaining rational solutions with the absolute free charges $ |Q'|<2 $.

In the next section, we present the general formalism for NSIs  and how these non-universal charges relate to the NSI parameters.

\section{Neutrino NSIs}
\label{sec:nsi}
New physics effects in the neutrino sector, such as couplings between neutrinos and unknown particles, can be described by a model independent four-fermion effective Lagrangian that corresponds to NSIs~\cite{Grossman:1995wx,Ohlsson:2012kf}. The NSI Lagrangian including Neutral Currents (NCs) can be parameterised in terms of the dimensionless NSI parameters $\varepsilon_{\alpha \beta}^{f X}$ as
\begin{equation}
\label{eq:NSILag}
\mathcal{L}_\mathrm{NSI}^\mathrm{NC}  = -2 \sqrt{2} G_F \varepsilon_{\alpha \beta}^{f X}  [\bar{f} \gamma^\mu P_X f][\bar{\nu}_\alpha \gamma_\mu P_L \nu_\beta] ,
\end{equation}
where $X$ is either $ L $ or $ R $,  $ f = u, \ d, \ e $ and $ G_F $ is the Fermi constant. Neutrino flavours are given by $  \alpha, \ \beta \ = e, \ \mu, \ \tau $. In the case of $ \alpha \neq \beta $, the NSI parameters imply flavour--violating new physics interactions in Eq.~(\ref{eq:NSILag}), while $ \alpha =\beta $  indicates flavour-conserving NSI terms. The former lead to zero-distance flavour-changing effects, which one can probe with the near detector of oscillation experiments. Both flavour-conserving and flavour-violating effects can lead to a modification of matter oscillations \cite{Wolfenstein:1977ue,Mikheyev:1985zog} to which IceCube is sensitive. Since gauge interactions are (nearly) flavour-diagonal, we concentrate on flavour-conserving interactions in what follows.

Considering the effective Lagrangian in  Eq.~(\ref{eq:NSILag}), we have a relation between the NSI parameters and the propagator of the mediator, $ \varepsilon_{\alpha \beta}^{f X} \propto \dfrac{1}{q^2 - M^2} $, where $ q $ and $ M $ are the mediator  momentum and mass, respectively. Matter oscillations arise from the interference of unperturbed propagation and gauge boson exchange in the forward direction,  thus the limit $q^{2} \rightarrow 0$ applies, so that the mass term dominates in the denominator even if the gauge boson is light. Therefore, an additional $ Z' $ boson  which satisfies the ATOMKI anomaly could provide a non-trivial contribution to the matter NSI parameters. Using  the interaction terms in Eq.~(\ref{eq:NeuCurLag}), it is possible to generate the $ Z'$-mediated effective NSI Lagrangian in Eq.~(\ref{eq:NSILag}), with corresponding NSI parameters 
\begin{equation}
\label{eq:NSIparZp}
\varepsilon_{\alpha \beta}^{f X} = \dfrac{1}{2\sqrt{2}G_F}\dfrac{C_L^{\alpha \beta}C_X^{ff}}{M_{Z'}^2}.
\end{equation}

The $ \varepsilon_{\alpha \beta}^{f X} $ are the effective couplings of neutrinos with fundamental fermions and affect neutrino propagation in matter. The relevant NSI effective couplings for neutrino propagation in a medium are their vector parts,  $ \varepsilon_{\alpha \beta}^{f V} =  \varepsilon_{\alpha \beta}^{f L} + \varepsilon_{\alpha \beta}^{f R}$, and the total strength of NSIs for a given medium has the form
\begin{equation}
\label{eq:epsilon}
\epsilon_{\alpha\beta} = \sum_f (\varepsilon_{\alpha\beta}^{f V})\dfrac{N_f}{N_e},
\end{equation}
where $ f = u, d, e $. Here $ N_f$ is the number density of the fermion $f$ in matter. Inside the Sun, $N_u/N_e \simeq 2N_d/N_e \simeq 1$ \cite{Serenelli:2009yc} while inside the Earth,
$N_u/N_e \simeq N_d/N_e \simeq 3 $ \cite{Lisi:1997yc}. Notice that the axial vector part of the current does not contribute and hence matter oscillations will not constrain it. In the presence of NSI couplings of neutrinos with the matter field $f$, the effective Hamiltonian is written as 
\begin{equation}
\label{eq:hamiltonian2}
H = \frac{1}{2E_\nu} U_{\rm PMNS} \diag (0, \Delta m_{21}^2, \Delta m_{31}^2) \, U_{\rm PMNS}^\dagger + 
V_{\rm CC} \diag (1, 0, 0) + V_{\rm CC} \,\epsilon_{\alpha\beta}~, 
\end{equation} 
where $ U_{\rm PMNS} $ is the vacuum Pontecorvo-Maki-Nakagawa-Sakata (PMNS) matrix while $ E_\nu $ and $ \Delta m_{ij}^2 \equiv \Delta m_{i}^2 -\Delta m_{j}^2$ are the neutrino energy and mass square differences, respectively. The second term describes the SM interactions in an unpolarised medium with the Wolfenstein matter potential $ V_{\rm CC} = \sqrt{2} G_F N_e $ \cite{Wolfenstein:1977ue}, where $ N_e $  is the local electron number density. The last term of Eq.~(\ref{eq:hamiltonian2}) is the NSI contribution, where the Hermitian matrix of the  NSI strength parameters $\epsilon_{\alpha \beta}$ shown in Eq.~(\ref{eq:epsilon}) can be written as 

\begin{equation}
\label{eq:epsilon2}
\epsilon_{\alpha \beta}=
\left(
\begin{array}{ccc}
\epsilon_{ee}  & \epsilon_{e\mu}  & \epsilon_{e\tau}  \\
\epsilon_{e\mu}^\ast &  \epsilon_{\mu\mu} & \epsilon_{\mu\tau}  \\
\epsilon_{e\tau}^\ast & \epsilon_{\mu\tau}^\ast  & \epsilon_{\tau\tau}  
\end{array}
\right)~. 
\end{equation}

The diagonal terms in Eq.~(\ref{eq:epsilon2}), if non-universal, lead to enhanced matter oscillations proportional to the difference in the diagonal NSI parameters, i.e.,  to $\epsilon_{\tau \tau} - \epsilon_{\mu \mu} $ and $\epsilon_{ee}-\epsilon_{\mu\mu}$.   Since any flavour-universal part gives just an unobservable common phase to the neutrinos,  one can subtract $\epsilon_{\mu \mu} $  from the diagonal in Eq.~(\ref{eq:epsilon2}), then the diagonal part of the matrix $\epsilon_{\alpha \beta}$ can be written as $ {\rm diag}(\epsilon_{ee}-\epsilon_{\mu \mu}, 0, \epsilon_{\tau \tau}-\epsilon_{\mu \mu}) $.

We approximate the hadrons to consist of their valence quarks\footnote{The $Z^{\prime}$, being uncoloured, does not see the gluonic sea. Since the momentum transfer is low, it cannot resolve the internal structure of the proton. Hence quark-antiquark pairs, having opposite charges (being dipole-like objects), will to first approximation look neutral and only the monopole charges of the valence quarks will be seen by the $Z^{\prime}$ boson.}, so we write the NSI parameters of the effective matter potential in terms of electron, proton and neutron NSI parameters as

\begin{equation}
\label{eq:epsilon3}
\epsilon_{\alpha\beta}^\oplus = \epsilon_{\alpha\beta}^{e V} +\epsilon_{\alpha\beta}^{p V }+Y_n^\oplus \epsilon_{\alpha\beta}^{n V},
\end{equation}
where $ \epsilon_{\alpha\beta}^{p V } = 2\epsilon_{\alpha\beta}^{u V } + \epsilon_{\alpha\beta}^{d V }$  ,  $ \epsilon_{\alpha\beta}^{n V } = 2\epsilon_{\alpha\beta}^{d V } + \epsilon_{\alpha\beta}^{u V }$ and $ Y_n^\oplus $ is the relative neutron-to-electron number density of the Earth, $ Y_n^\oplus \equiv N_n / N_e \approx 1.051$ \cite{Esteban:2018ppq}. Finally, one can obtain the NSI matrix  in the Hamiltonian with new definitions as 

\begin{equation}
\label{eq:epsilon4}
\epsilon_{\alpha \beta}^{\oplus}=
\left(
\begin{array}{ccc}
\epsilon_{ee}^\oplus-\epsilon_{\mu\mu}^\oplus   & \epsilon_{e\mu}^\oplus  & \epsilon_{e\tau}^\oplus  \\
\epsilon_{e\mu}^{\oplus \ast}&  0 & \epsilon_{\mu\tau}^\oplus  \\
\epsilon_{e\tau}^{\oplus \ast} & \epsilon_{\mu\tau}^{\oplus \ast} & \epsilon_{\tau\tau}^\oplus -\epsilon_{\mu\mu}^\oplus 
\end{array}
\right)~. 
\end{equation}

Notice again that there is no flavour violation (non-diagonal terms) in the theoretical model of  this work.

\section{Constraints on NSIs from IceCube}
\label{sec:external}

The previous generic parameterisation of the strength of NSIs as shown in Eq.~(\ref{eq:epsilon4}) has been used by the IceCube collaboration in Ref.~\cite{IceCubeCollaboration:2021euf} to constrain the parameters $\epsilon_{ee}^{\oplus }-\epsilon_{\mu\mu}^{\oplus }$, $\epsilon_{\tau \tau}^{\oplus } - \epsilon_{\mu \mu}^{\oplus } $, $|\epsilon_{e\mu}^{\oplus }|$ $|\epsilon_{e\tau}^{\oplus }|$ and $|\epsilon_{\mu\tau}^{\oplus }|$ using a pure sample of atmospheric neutrinos (and antineutrinos) of all flavours with energies between $5.6$~GeV and $100$~GeV. The use of atmospheric neutrinos allows to sample a wide range of oscillation baselines, from a few tens of kilometres, for downgoing neutrinos produced ``above'' the detector that only cross the atmosphere, to the whole diameter of the Earth, $1.3\times 10^4$ km, for upgoing neutrinos produced at the antipodes of the detector. Matter effects can thus be expected for neutrinos arriving to the detector from below the horizon, while the atmosphere is too thin to induce any matter effects on the neutrino flux arriving to the detector from above.

Comparing the  measured flavour composition of the neutrino flux at the detector as a function of energy and baseline with the expected corresponding flux under standard oscillations, strong limits on the NSI parameters can be set. Note that these constraints were obtained by allowing one of the parameters to be non-zero at a time. We do not consider flavour-violating terms in our study, so we only use the IceCube limits on flavour-diagonal interactions, shown in Tab.~\ref{tab:constraints}, in order to put constraints on the model described in Sec.~\ref{sec:model}.

\section{Constraints on the $U(1)'$ model from other experiments}
\label{sec:numeric}

To define the parameter space of our model we have used the SPheno~\cite{Porod:2003um,Porod:2011nf,Braathen:2017izn} and SARAH 4.14.3~\cite{Staub:2013tta,Staub:2015kfa} codes. The scanning of the parameter space was performed using the Metropolis-Hastings algorithm, within the ranges specified in Tab.~\ref{paramSP}.

\begin{table}[t!]
	\centering
	\begin{tabular}{c|c||c|c}
		\hline
		Parameter  & Scanned range & Parameter      & Scanned range \\
		\hline
		$g'$ & $[10^{-5}, 5\times 10^{-5}]$      & $\lambda$ & $[0.125, 0.132]$ \\
		$\tilde{g}$        & $[-10^{-3}, 10^{-3}]$ & ${\lambda}_{\chi}$ & $[10^{-5}, 10^{-1}]$ \\
		$v'$ & $[0.1, 1]$ TeV  &  $\kappa$  & $[10^{-6}, 10^{-2}]$ \\
		$  Q'_{Q_1}, Q'_{Q_3},Q'_{L_1}$ & $[-2, 2]$ &$Q'_{\nu_1} $& $0$ \\
		\hline
	\end{tabular}
	\caption{Scanned parameter space of the model.}
	\label{paramSP}
\end{table}

Before applying the NSI constraints from IceCube \cite{IceCubeCollaboration:2021euf} to our $U(1)'$ model, some important experimental outcomes have been implemented to our results from the  parameter space scan.
We first require the Higgs boson mass to be within $3$~GeV of its observed value of 125 GeV and implement constraints on the Branching Ratios (BRs) of rare $B$-decays, specifically $ {\rm BR}(B \rightarrow X_{s} \gamma) $, $ {\rm BR}(B_s \rightarrow \mu^+ \mu^-) $ and $ {\rm BR}(B_u\rightarrow\tau \nu_{\tau}) $ \cite{HFLAV:2012imy,LHCb:2012skj,HFLAV:2010pgm}. We have also bounded the value of the $ Z-Z' $ mixing parameter $\theta'$ (see Eq.~(\ref{ThetaPrime})) to be less than a few times $ 10^{-3} $ as a result of EW  Precision Tests (EWPTs)~\cite{DELPHI:1994ufk,Erler:2009jh}. 
 
 In the following part of the numerical analyses, we constrain the parameter space to satisfy the current experimental bounds on AMM results for $ (g-2)_{e} $, $ (g-2)_{\mu} $ \cite{Muong-2:2023cdq,Morel:2020dww}, the ATOMKI anomaly \cite{Barducci:2022lqd}, the electron beam dump experiment NA64 \cite{NA64:2019auh}, which probes the $Z'$ couplings to electrons, $ C^2_{e,V}+ C^2_{e,A} $, in any possible production and detection via $ e^- + Z → e^- + Z'[→ e^+ e^-] $ and  neutrino scattering from the TEXONO experiment \cite{TEXONO:2009knm}.  
 
{In the process of  our Benchmark Point (BP) selection, we have also applied additional experimental limits on parity-violating Moller scattering \cite{SLACE158:2005uay,Kahn:2016vjr}, atomic parity violation \cite{Porsev:2009pr,Arcadi:2019uif} and Coherent Elastic neutrino-Nucleus Scattering (CEvNS) \cite{Denton:2023gat,AristizabalSierra:2022axl,AtzoriCorona:2022moj,Coloma:2022avw} as $  |C_{e,V} \times C_{e,A}| \lesssim 10^{-8} $,  $ |C_{e,A}| \left| \frac{188}{399}  C_{u,V}  + \frac{211}{399} C_{d,V }\right| \lesssim 1.8\times10^{-12}  $ and $ \sqrt{\left| C_n C_{\nu_e}\right|} \lesssim 13.6 \times 10^{-5}$, respectively, Here, $ C_n $ is the $Z'$-neutron coupling defined as $ C_n =  C_{u,V}  + 2 C_{d,V}$. The experimental constraints are summarised in Tab.~\ref{tab:constraints}.  }
 
\begin{table}[t!]
\centering
\begin{tabular}{c|c|c|c}
\hline
Observable & Constraint & Tolerance & Reference(s) \\		
\hline
$m_h$ & 122 GeV -- 128 GeV & ~ & ~ \\
${\rm BR}(B_s \rightarrow \mu^{+} \mu^{-} )$ &
$0.8 \times 10^{-9}$  --
$6.2 \times 10^{-9}$ &
$2\sigma$ & 
\cite{LHCb:2012skj} \\
${\rm BR}(B \rightarrow X_{s} \gamma)$ &
$2.99 \times 10^{-4}$ --
$3.87 \times 10^{-4}$ &
$2\sigma$ & 
\cite{HFLAV:2012imy} \\
$\frac{{\rm BR}(B_u\rightarrow\tau \nu_{\tau})}{{\rm BR}(B_u\rightarrow \tau \nu_{\tau})_{\rm SM}}$ &
0.15  --
2.41 &
$3\sigma$ & 
\cite{HFLAV:2010pgm} \\
$\Delta a_e^{\text{Rb}}$ &
$(4.8 \pm 9.0) \times 10^{-13}$ &
$3\sigma$ & 
\cite{Morel:2020dww} \\
$\Delta a_{\mu}$ &
$(2.45 \pm 1.47)\times 10^{-9}$ &
$3\sigma$ & 
\cite{Muong-2:2023cdq} \\
$\epsilon_{ee}^{\oplus }-\epsilon_{\mu\mu}^{\oplus }$ &
$[-2.26, -1.27] \cup [-0.74, 0.32]$ &
~ &
\cite{IceCubeCollaboration:2021euf} \\
$\epsilon_{\tau \tau}^{\oplus }-\epsilon_{\mu\mu}^{\oplus } $ &
$[-0.041,0.042]$ &
~ &
\cite{IceCubeCollaboration:2021euf} \\
$ \sqrt{C^2_{e,V}+ C^2_{e,A}} $ &
$\gsim 3.6 \times 10^{-5} \times \sqrt{BR(Z' \to e^+ e^-)}$ &
~ &
\cite{NA64:2019auh} \\
$ \sqrt{C_{e,V} \times C_{\nu_e}} $ &
$\lesssim 3 \times 10^{-4} $ &
~ &
\cite{TEXONO:2009knm} \\
$ |C_{e,V} \times C_{e,A}| $ &
$\lesssim 10^{-8}  $ &
~ &
\cite{SLACE158:2005uay,Kahn:2016vjr} \\
$ |C_{e,A}| \left| \frac{188}{399}  C_{u,V}  + \frac{211}{399} C_{d,V }\right|  $ &
$\lesssim 1.8 \times 10^{-12}  $ &
~ &
\cite{Porsev:2009pr,Arcadi:2019uif}  \\
$ \sqrt{|C_{n} \times C_{\nu_e}|} $ &
$\lesssim 13.6 \times 10^{-5} $ &
~ &
\cite{Denton:2023gat,AristizabalSierra:2022axl,AtzoriCorona:2022moj,Coloma:2022avw}\\
\hline
\end{tabular}
\caption{Summary of the experimental constraints used.}
\label{tab:constraints}
\end{table}

\section{Results}
\label{sec:results}

\begin{figure}[t!]
	\centering
	\includegraphics[scale=0.52]{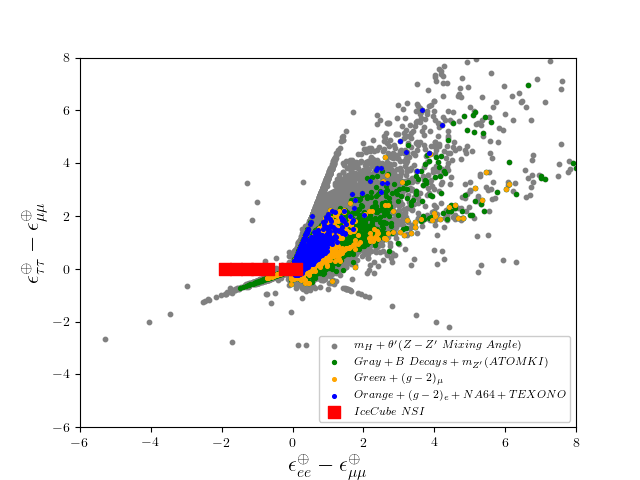}
	\caption{The distribution of diagonal NSI parameters  shown in Eq.~(\ref{eq:epsilon4}). All points are consistent with Higgs mass bounds and the $ Z-Z' $ mixing satisfying EWPTs. Green points are a subset of the grey ones as they also satisfy constraints on  $B$-decays and $ Z' $ mass around 17 MeV.  Yellow points are a subset of the green ones as they are also compatible with the current experimental bounds of $ (g-2)_{\mu} $ while blue points, that are a subset of the yellow ones, also  satisfy the experimental limits from $ (g-2)_{e} $, NA64 and TEXONO. The red squares inside the blue points represent the region allowed by IceCube results in Table \ref{tab:constraints}.}
	\label{fig:NSI_param}
\end{figure}
In this section, we present the numerical analysis in the light of the experimental constraints from the previous section. First, let us focus on the diagonal NSI parameters shown in Eq.~(\ref{eq:epsilon4}). Fig.~\ref{fig:NSI_param} depicts the distribution of these parameters after scanning the parameter space. All points are consistent with Higgs mass bounds and the $ Z-Z' $ mixing satisfying EWPTs. Green points are a subset of the grey ones as they also satisfy constraints on  $B$-decays and $ Z' $ mass around 17 MeV.  Yellow points are a subset of the green ones as they are also compatible with the current experimental bounds of $ (g-2)_{\mu} $ while blue points, that are a subset of the yellow ones, also  satisfy the experimental limits from $ (g-2)_{e} $, NA64 and TEXONO. The red painted area represents the region allowed by IceCube results. As can be seen from the figure, most of our solutions are ruled out by the IceCube bounds on NSI parameters.

\begin{figure}[t!]
	\centering
	\includegraphics[scale=0.45]{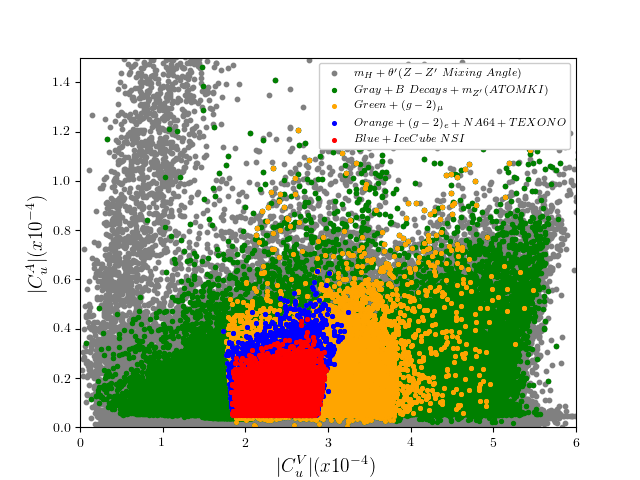}
	\includegraphics[scale=0.45]{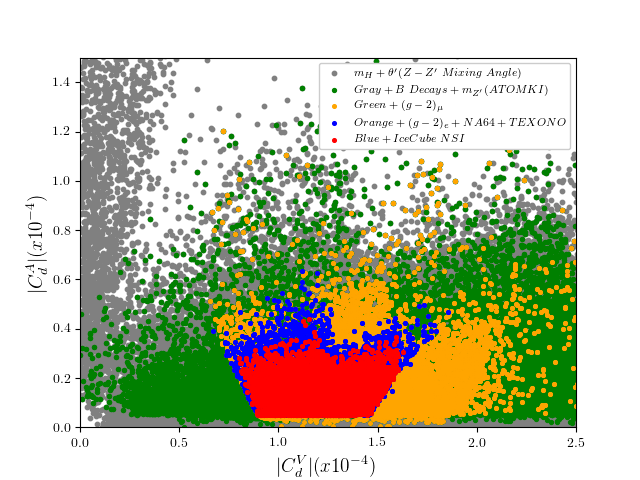}
	\includegraphics[scale=0.45]{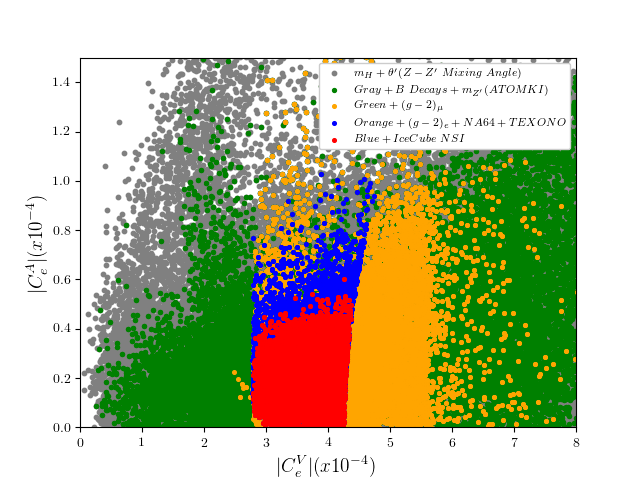}
	\caption{The allowed regions for the vector and axial-vector $ Z' $ couplings with up-quarks (top left), down-quarks (top right), electron (bottom). The colour convention is the same as in Fig.~\ref{fig:NSI_param} while  additional red points are a subset of the blue ones as they also satisfy the NSI parameters allowed by IceCube results.}
	\label{fig:ZpNeuCoup}
\end{figure}

In the light of  these strict bounds, we will show the constraints on $ Z' $ couplings. In Fig.~\ref{fig:ZpNeuCoup}, we represent the allowed vector and axial-vector $ Z' $ couplings with up-quarks (top left), down-quarks (top right) and electron (bottom). The colour convention is the same as in Fig.~\ref{fig:NSI_param} while  additional red points are a subset of the blue ones as they also satisfy the NSI parameters allowed by IceCube results. According to our results, the vector  couplings between $ Z' $ and $ u, d, e $, the fundamental particles in the  medium, tend to be in the small interval of  $\mathcal{O}(10^{-4}) $ order while the axial-couplings should be at the order of $\mathcal{O}(10^{-5})$. Here, the vertical limitations on the couplings mainly arise from the experimental bound on the AMM of the electron in Tab. \ref{tab:constraints} which is used for blue and then red points. As can be seen from Eq. (\ref{gm2s2}),  $ C_{e,V}$ has lower and upper limits for  $ C_{e, A}\simeq   \mathcal{O}(10^{-5})$.  These limits also impacts the couplings of the $ u $ and $ d $ quarks with $ Z' $ through the charge relations  in Eq. \ref{eq:charges}. Since the effective NSI couplings are related to the vector parts of the NSI parameters shown in Eq. (\ref{eq:epsilon}),  the vector couplings are strongly bounded by IceCube  constraints on NSIs. 

\begin{figure}[t!]
	\centering
	\includegraphics[scale=0.46]{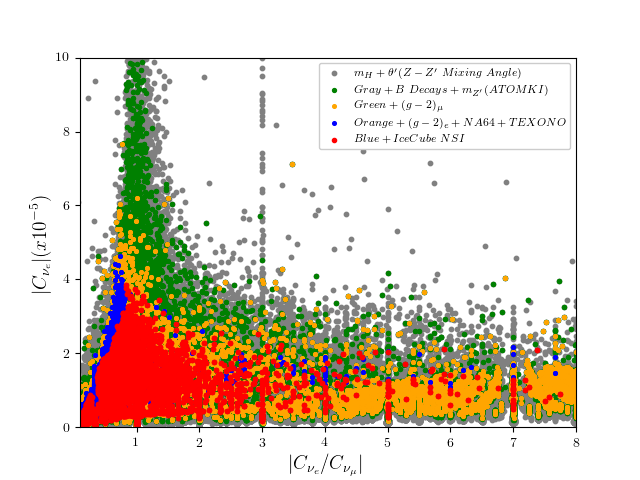}
	\includegraphics[scale=0.46]{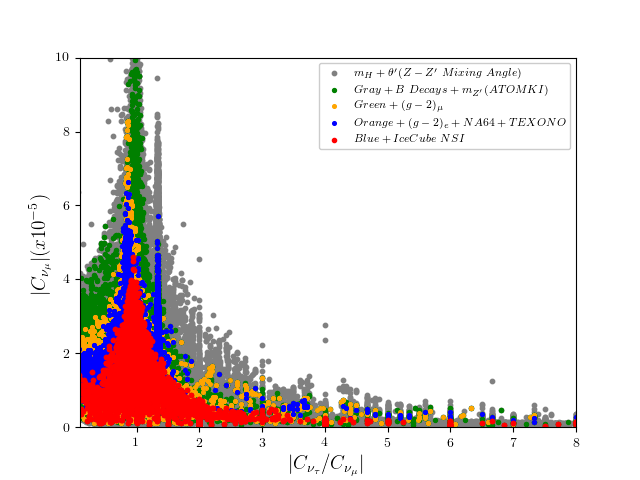}
	\includegraphics[scale=0.46]{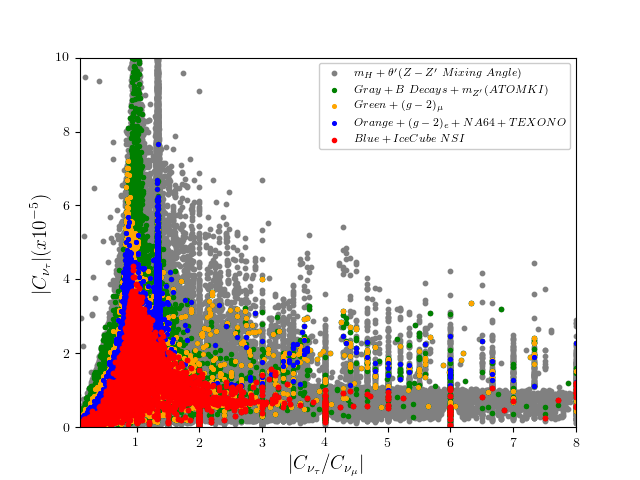}
	\caption{The allowed regions for the $ Z' $ couplings with $ \nu_e, \nu_\mu $ (top) and $ \nu_\tau $  (bottom)  versus their ratios to each other. Here, the colour convention is the same as in Fig.~\ref{fig:ZpNeuCoup}.}
	\label{fig:ZpNeuCoup2}
\end{figure}

Furthermore, Fig.~\ref{fig:ZpNeuCoup2} presents the  distributions  of electron neutrino (top left), muon neutrino (top right) and tau neutrino (bottom) in terms of their ratios to each other.  As can be seen from the panels in the figure, when all constraints are applied, the $ Z' $ couplings with all neutrinos are restricted to be smaller than of  $\mathcal{O}(5\times10^{-5})$.  Since, in  principle, the diagonal NSI terms in Eq.~(\ref{eq:epsilon4}) arise from the non-universality in the lepton sector, it is not hard to guess that the major impact of the IceCube results will be on non-universality of the $ Z' $-lepton couplings. When we also look at the ratios of the $ Z' $ couplings with each neutrino flavour, it can be easily seen that the number of solutions for  $ |C_{\nu_\tau}/C_{\nu_\mu}| $ that satisfy IceCube constraints condense around $ 1 $. It means that those  for $ \mu $ and $\tau$ neutrinos should be universal, $ |C_{\nu_\mu}/C_{\nu_\tau}| \approx 1$, because of the  small values of $ \epsilon_{\tau\tau}^\oplus -\epsilon_{\mu\mu}^\oplus  $,  according to  the NSI bounds from IceCube. In contrast, the ratio for electron neutrino coupling such as $ |C_{\nu_e}/C_{\nu_\mu}| $, can be seen as larger values in much more solutions, for the reason that the $ \epsilon_{ee}^\oplus -\epsilon_{\mu\mu}^\oplus  $ parameter gets looser bounds from IceCube data. There are also certain other discrete values of coupling ratios that seem to be preferred. These are cases, where the Higgs charge $Q^{\prime}_{H}$ is either exactly or close to zero (see eq. (\ref{eq:charges})). In that case the Higgs charge gives a smaller contribution to the mixing angle $\theta^{\prime}$ (eq. (\ref{ThetaPrime})) than otherwise and hence there will be a lot more points that are acceptable based on the $Z$--$Z^{\prime}$ mixing.

\begin{figure}[t!]
	\centering
	\includegraphics[scale=0.46]{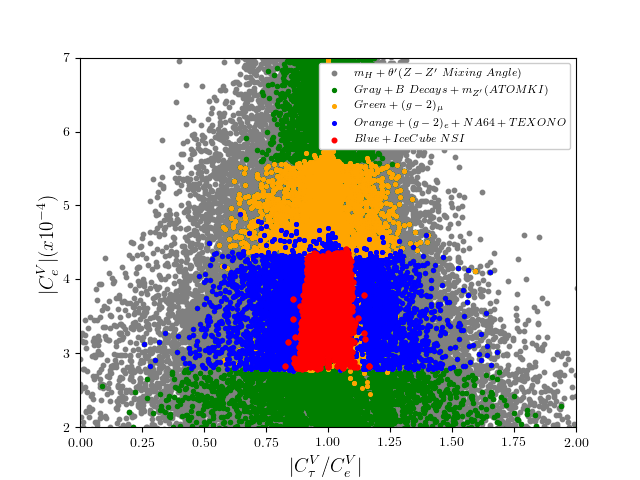}
	\includegraphics[scale=0.46]{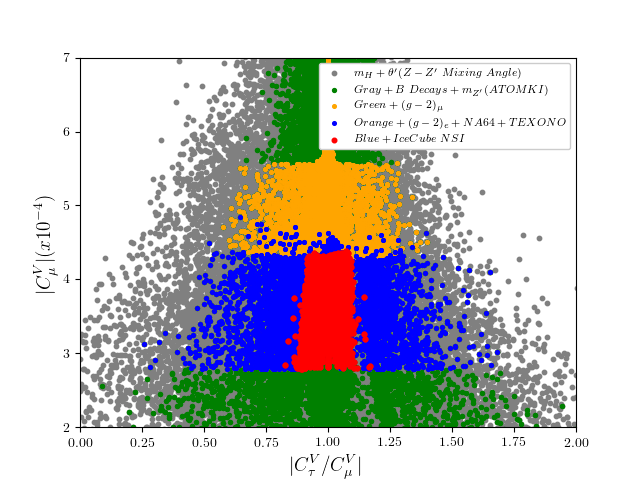}
	\includegraphics[scale=0.46]{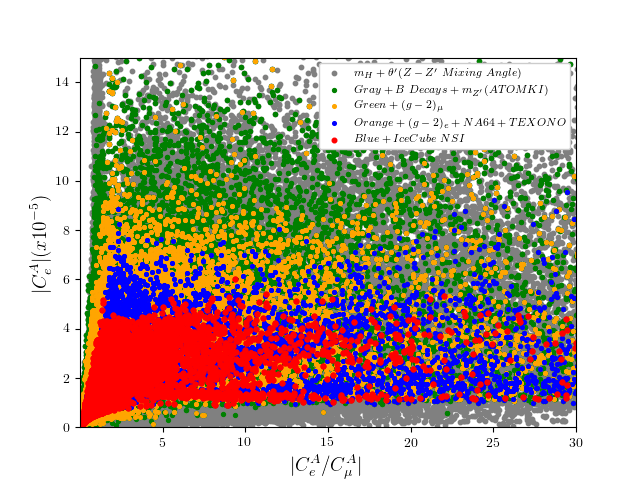}
	\includegraphics[scale=0.46]{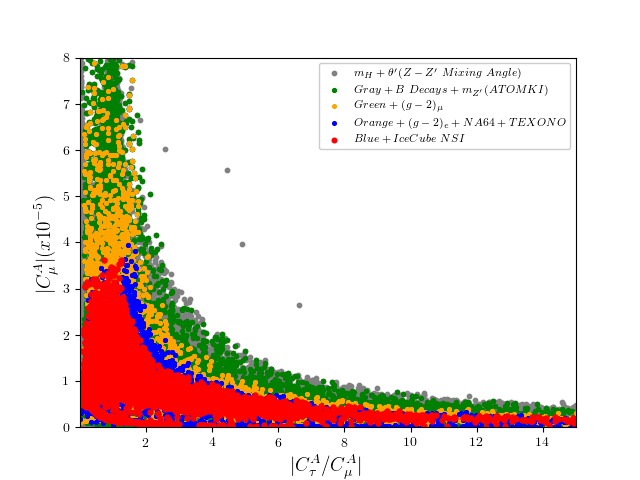}
	\caption{The allowed regions for the ratios of $ Z' $ couplings with $ e, \mu $ and $ \tau $ leptons (top) and neutrinos (bottom). Here, the colour convention is the same as in Fig.~\ref{fig:ZpNeuCoup}.}
	\label{fig:ZpNeuCoup3}
\end{figure}

Fig.~\ref{fig:ZpNeuCoup3} indicates the allowed regions for the ratios of $ Z' $ couplings with charged leptons (top) and neutrinos (bottom). Here, the colour convention is the same as in Fig.~\ref{fig:ZpNeuCoup}. As we expected, the ratios of vector couplings are mostly bounded. It is important to note that, after applying  the  NSI constraints from IceCube as well, the vector couplings between  $ Z' $ and charged leptons should be of the same magnitude, $ |C^V_{\tau}/C^V_{e}| \approx |C^V_{\tau}/C^V_{\mu}| \approx 1$, that can be deemed  universal, while the axial-vector $ Z' $ couplings for each charged lepton, not constrained by IceCube data, allow significant  non-universality, as $ 0 < |C^A_{e}/C^A_{\mu}|< 30$ and $ 0 < |C^A_{\tau}/C^A_{\mu}|< 8$. Therefore, one can expect that  possible signatures for such theoretical frameworks could be explored in  Lepton Flavour Violation (LFV) processes which are more sensitive to the new boson axial couplings than its vector couplings. 

\begin{figure}[t!]
	\centering
	\includegraphics[scale=0.46]{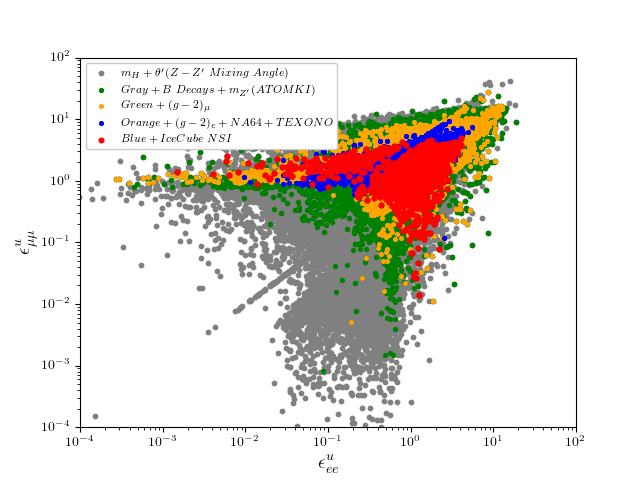}
	\includegraphics[scale=0.46]{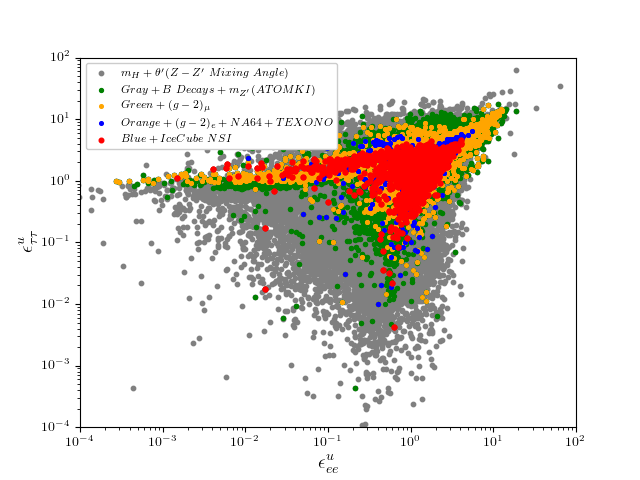}
	\includegraphics[scale=0.46]{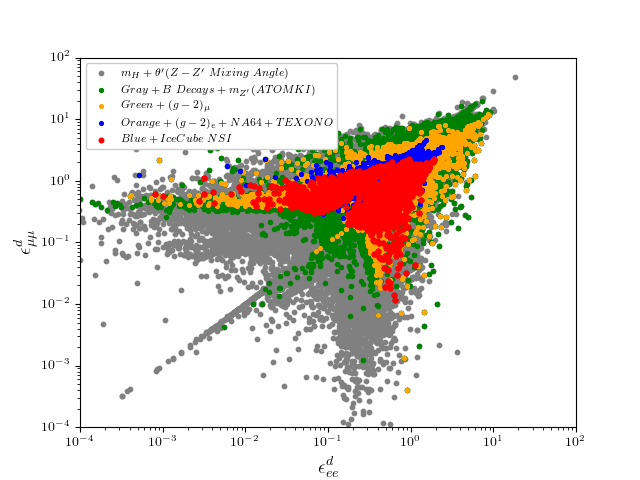}
	\includegraphics[scale=0.46]{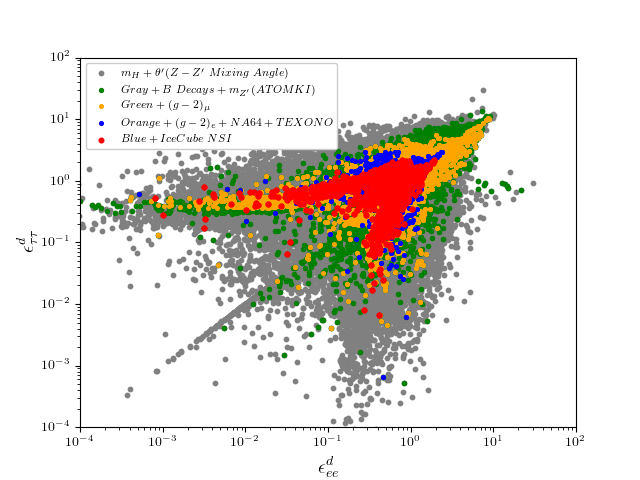}
	\includegraphics[scale=0.46]{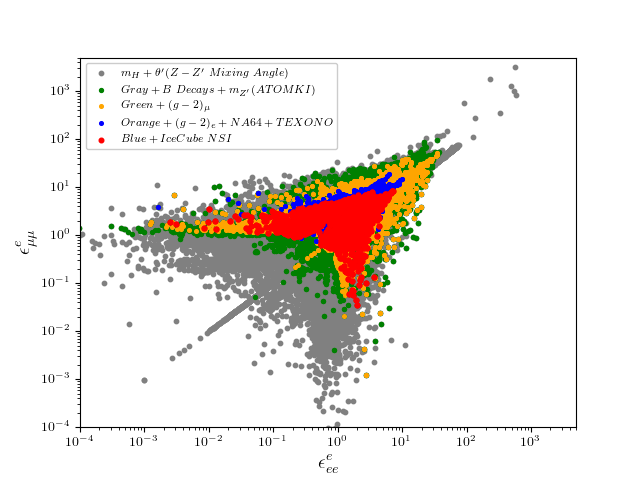}
	\includegraphics[scale=0.46]{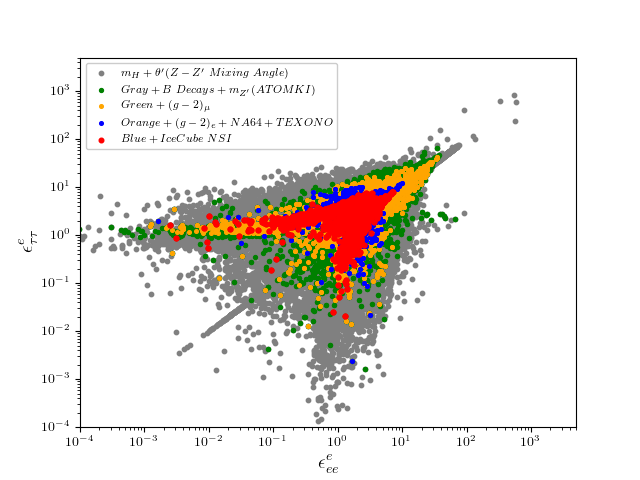}
	\caption{The allowed regions  for NSI effective couplings of neutrinos with up-quarks (top), down-quarks  (middle) and electrons (bottom) via $ Z' $ mediation. The colour convention is the same as in Fig.~\ref{fig:ZpNeuCoup}.}
	\label{fig:EffZpCoup}

\end{figure}

Before closing, we investigate how the IceCube bounds impact the relevant NSI effective couplings for neutrino propagation. Fig.~\ref{fig:EffZpCoup} displays the allowed regions  for NSI effective couplings of neutrinos with up-quarks (top), down-quarks  (middle) and electrons (bottom) via $ Z' $ mediation. The colour convention is the same as in Fig.~\ref{fig:ZpNeuCoup}. As can be seen from the graphs, in the light of the IceCube results, all effective couplings are restricted to the region with $ \epsilon \lesssim 10$. Considering the solutions which satisfy all experimental constraints except for NSI bounds (blue points), it is clear that the effective couplings, via  light vector boson mediation between neutrinos and the components of the atoms in the medium, are  limited by IceCube results.

To finalise our discussion about these results, we display four BPs: in fact, Tab.~\ref{tab:benchmark1} displays four solutions which are selected to be consistent with all experimental constraints applied in our analyses as well as  NSI results from IceCube.

\begin{table}[t!]
	\centering\hspace*{-1.1truecm}
	\scalebox{0.7}{
		\begin{tabular}{|c|c|c|c|c|}
			\hline
			Parameters &  BP1 & BP2 & BP3 & BP4 \\ \hline
			$g^\prime$ &$ 1.16\times 10^{-5}$&$ 1.45 \times 10^{-5}$&$ 2.31 \times 10^{-5}$&$ 1.20 \times 10^{-5}$\\
			$\Tilde{g}$ &$-4.97\times 10^{-4}$&$ -4.69 \times 10^{-4}$&$ -5.10 \times 10^{-4}$&$ -5.32 \times 10^{-4}$\\
			$v^{\prime}$ &746&588&369&691\\
			$\lambda_\chi$&0.007&0.003&0.006&0.009\\
			$\kappa$ &0.0079&0.0123&0.0087&0.0021\\ \hline
			$(m_{H_1},\ m_{H_2})$ & 
			(87.46,\,124.34) &
			(45.30,\,124.29) &
			(40.66,\,125.35) &
			(93.86,\,125.98)\\
			$m_{Z^\prime}$ & 
			0.0173&
			0.0172 &
			0.0171 &
			0.0166\\
			$(C^V_{u},C^A_{u})$ & 
			($ -2.45 \times 10^{-4}$,\,$ 1.41\times 10^{-5}$) &
			($ -2.51\times 10^{-4}$,\,$ -1.35\times 10^{-5}$) &
			($ -2.67 \times 10^{-4}$,\,$ -1.12 \times 10^{-5}$) &
			($ -2.78\times 10^{-4}$,\,$ -1.17 \times 10^{-5}$) \\
			$(C^V_{d},C^A_{d})$ & 
			($ 1.29\times 10^{-4}$,\,$ -1.41\times 10^{-5}$) &
			($ 1.18 \times 10^{-4}$,\,$ 1.34\times 10^{-5}$) &
			($ 1.27 \times 10^{-4}$,\,$ 1.12\times 10^{-5}$) &
			($ 1.33 \times 10^{-4}$,\,$ 1.17\times 10^{-5}$) \\
			$(C^V_{e},C^A_{e})$& 
			($ 3.61\times 10^{-4}$,\,$ 0 $) &
			($ 3.83 \times 10^{-4}$,\,$ 0 $) &
			($ 4.06 \times 10^{-4}$,\,$ 0 $) &
			($ 4.23 \times 10^{-4}$,\,$ 0 $) \\
			$(C_{\nu_e},C_{\nu_\mu})$& 
			($ -1.47 \times 10^{-5}$,\,$ -1.36 \times 10^{-5}$) &
			($ 1.42 \times 10^{-5}$,\,$ 1.27  \times 10^{-5}$) &
			($ 1.01 \times 10^{-5}$,\,$ 1.23 \times 10^{-5}$) &
			($ 1.05\times 10^{-5}$,\,$ 1.29 \times 10^{-5}$) \\
			$(\epsilon^u_{ee},\epsilon^u_{\mu \mu},\epsilon^u_{\tau \tau})$& 
			$(1.45,\,1.33,\,1.05)$ &
			$(-1.46,\,-1.31,\,-0.94)$ &
			$(-1.11,\,-1.37,\,-2.01)$ &
			$(-1.27,\,-1.56,\,-2.29)$\\
			$(\epsilon^d_{ee},\epsilon^d_{\mu \mu},\epsilon^d_{\tau \tau})$& 
			$(-0.76,\,-0.71,\,-0.55)$ &
			$(0.69,\,0.62,\,0.44)$ &
			$(0.53,\,0.65,\,0.96)$ &
			$(0.61,\,0.75,\,1.10)$\\
			$(\epsilon^e_{ee},\epsilon^e_{\mu \mu},\epsilon^e_{\tau \tau})$& 
			$(-2.13,\,-1.96,\,-1.54)$ &
			$(2.24,\,2.01,\,1.43)$ &
			$(1.69,\,2.08,\,3.06)$ &
			$(1.93,\,2.38,\,3.49)$\\
			$ \epsilon_{ee}^\oplus -\epsilon_{\mu\mu}^\oplus  $ & $-0.007$&$-0.008$&0.013&0.011\\
			$ \epsilon_{\tau\tau}^\oplus -\epsilon_{\mu\mu}^\oplus  $ & 0.017 &0.021 &$-0.028$&$-0.032$\\ \hline
			$  Q'_{u_1}  $& 2.0 &$-$1.2&$-$0.4&$-$0.8\\
			$  Q'_{u_2}  $& 2.0&$-$1.2&$-$0.4&$-$0.8\\
			$  Q'_{u_3}  $&1.6 &$-$0.8&$-$0.8&$-$1.61\\
			$  Q'_{d_1}  $& $-$1.0&0.6&0.2&0.4\\
			$  Q'_{d_2}  $&$-$1.0&0.6&0.2&0.4\\
			$  Q'_{d_3}  $&$-$1.0&0.6&0.2&0.4\\
			$  Q'_{Q_1}  $& 0.5&$-$0.3&$-$0.1&$-$0.2\\
			$  Q'_{Q_2}  $& 0.5 &$-$0.3&$-$0.1&$-$0.2\\
			$  Q'_{Q_3}  $& 0.3&$-$0.1&$-$0.3&$-$0.6\\
			$  Q'_{e_1}  $&$-$2.9&1.7&0.7 &1.4\\
			$  Q'_{e_2}  $& $-$3.1&1.9&0.5&1.0\\
			$  Q'_{e_3}  $&$-$2.2 &1.0&1.4 &2.8\\
			$  Q'_{L_1}  $& $-$1.6 &1.0&0.2&0.4\\
			$  Q'_{L_2}  $& $-$1.4 &0.8&0.4&0.8\\
			$  Q'_{L_3}  $& $-$0.9 &0.3&0.9&1.8\\
			$  Q'_{\nu_1}  $& 0 &0&0&0\\
			$  Q'_{\nu_2}  $& 0 &0&0&0\\
			$  Q'_{\nu_3}  $& $-$0.4 &0.4&$-$0.4&$-$0.8\\
			$  Q'_{H}      $& 1.3&$-$0.7&$-$0.5&$-$1\\
			\hline
	\end{tabular}}
	\caption{The BPs which are selected to be consistent with all experimental constraints as well as  NSI results from IceCube. All masses  are given in GeV.}
	\label{tab:benchmark1}
\end{table}

\section{Conclusions}
\label{sec:conc}
We have described a rather simple theoretical framework that relies
on a $U(1)'$ extension of the SM with non-anomalous and flavour-dependent charges allowing for vector and axial-vector couplings of a new $ Z' $ state to nucleons. The $Z'$ has a mass of $O(10)$ MeV and emerges from the spontaneous breaking of the new gauge  group, 
allowing it to be a possible explanation  of  the $X17$ anomaly. However,  in order to comply with experimental bounds on flavour-violation in the quark sector, we have  imposed that the first two quark  generations are  flavour-universal under this  $U(1)'$ gauge group  while the corresponding 
charges  of the  lepton sector are  left as fully non-universal. As a consequence, couplings of the $Z'$ state with all  light neutrinos are present in the model and may manifest themselves in NSIs of neutrinos  affecting neutrino flavour ratios  in matter.

We have constrained this theoretical construction with data from the ATOMKI collaboration and other low energy  experiments, such as NA64 searches for $Z'$, data on $(g-2)_{e,\mu}$, constraints from CEvNS, Moller scattering and atomic parity violation, and finally with NSI data from the IceCube neutrino experiment (complementing earlier data from TEXONO). 
 
IceCube data constrain the vector parts of the $Z^{\prime}$ interactions with leptons to be nearly flavour-universal while they give no constraints on the universality of the axial vector part. In the neutrino sector the constraints on $\mu$ -- $\tau$ universality are strong while $e$ -- $\mu$ universality is somewhat less constrained by the IceCube results. The results also exclude a scenario where the $Z'$ couples only to electrons. This means that almost flavour-universal couplings with charged leptons are possible, facilitating searches in leptonic channels at colliders and low-energy experiments.

We have in the end found that there are sizeable regions of parameter space in this theoretical framework that can accommodate all such constraints. We have also defined four benchmark points that can be used in further phenomenological investigation of the model described.

\subsection*{Acknowledgements}
 SM is supported in part through the NExT Institute and the STFC Consolidated Grant No. ST/L000296/1. The work of YH is supported by Balikesir University Scientific  Research Projects with Grant No. BAP-2022/083. YH thanks the other authors for their hospitality during a visit to Uppsala, which initiated this work, under the auspices of the Erasmus$+$ Staff Mobility for Training. CPH is supported by grant no. 2021-04759 from the Swedish Research Council. HW is supported by the Carl Trygger Foundation under grant no. CTS18:164 and the Ruth and Nils-Erik Stenb\"acks Foundation.

\end{document}